\documentclass[aps,prb,preprint,groupedaddress,showpacs]{revtex4}
\usepackage{epsfig}
\usepackage{subfigure}
\usepackage{amssymb}
\usepackage{graphicx}
\usepackage{amsmath}
\usepackage{natbib}
\newcommand{\mb}{\boldsymbol}
\begin{document}

\title{Random walk approach to spin dynamics in a two-dimensional electron gas with spin-orbit coupling}

\author{Luyi Yang}
\author{J. Orenstein}
\author{Dung-Hai Lee}
\affiliation{Department of Physics, University of California,
Berkeley, California 94720, USA} \affiliation{Materials Science
Division, Lawrence Berkeley National Laboratory, Berkeley,
California 94720, USA}


\begin{abstract}
We introduce and solve a semi-classical random walk (RW) model that
describes the dynamics of spin polarization waves in zinc-blende
semiconductor quantum wells. We derive the dispersion relations for
these waves, including the Rashba, linear and cubic Dresselhaus
spin-orbit interactions, as well as the effects of an electric field
applied parallel to the spin polarization wave vector. In agreement
with calculations based on quantum kinetic theory [P. Kleinert and
V. V. Bryksin, Phys. Rev. B \textbf{76}, 205326 (2007)], the RW
approach predicts that spin waves acquire a phase velocity in the
presence of the field that crosses zero at a nonzero wave vector,
$q_0$. In addition, we show that the spin-wave decay rate is
independent of field at $q_0$ but increases as $(q-q_0)^2$ for
$q\neq q_0$. These predictions can be tested experimentally by
suitable transient spin grating experiments.
\end{abstract}


\maketitle

\section{Introduction\label{sec:intro}}
Spin-orbit (SO) coupled two-dimensional electron systems are of
great interest, both as model systems and as the active component of
devices that control electron spin with electric fields.\cite{Dietl}
Unfortunately, the potential of the SO interaction to control
electron spin comes with a price - the SO terms in the Hamiltonian
break SU(2) spin symmetry. The violation of SU(2) means that
electron spin polarization is not conserved, decaying instead with a
characteristic spin memory time $\tau_s$. The mechanism by which SO
coupling leads to spin memory loss has been intensively investigated
in two-dimensional electron gases (2DEGs) in semiconductor quantum
wells (QWs), as described in recent reviews.\cite{Fabian, Wu} In
GaAs QWs and related systems, breaking of inversion symmetry allows
SO coupling that is linear in the electron wave vector
$\mb{k}$.\cite{Rashba1, Rashba2, Rashba3} The SO terms in the
Hamiltonian can be viewed as effective magnetic fields that act only
on the electron spin, with magnitude and direction that vary with
$\mb{k}$. The loss of spin memory in the effective magnetic field,
$\mb{b}(\mb{k})$, takes place through the D'yakonov-Perel' (DP)
mechanism.\cite{DP1, DP2, DP3, DP4} In this process the electron
spin precesses during its ballistic motion between collisions; each
time it is scattered $\mb{b}(\mb{k})$ and consequently the
precession vector, $\mb{\Omega}(\mb{k})$, change. The net result is
exponential decay of spin polarization at a rate approximately equal
to $\Omega^2\tau$, where $\tau$ is the mean time between collisions.

There exist two distinct contributions to $\mb{b}(\mb{k})$, the
Rashba term\cite{Rashba2, Rashba3} arising from asymmetry of the
confining potential and the Dresselhaus term\cite{Dresselhaus}
originating in the intrinsic inversion asymmetry of the GaAs crystal
structure. A prescription for lengthening spin lifetime in QWs of
III-V semiconductors by tuning the Rashba coupling strength
($\alpha$) to equal the linear Dresselhaus coupling ($\beta_1$) was
proposed by Schliemann \emph{et al.}\cite{SL}  Recently it was
recognized that this mechanism amounts to a restoration of SU(2)
symmetry even in the presence of anisotropic SO
interactions.\cite{SU2} The main purpose of this paper is to assess
theoretically to what extent tuning SO interactions can be expected
to increase the distance over which electron spin polarization can
propagate without decay.

The potential to extend the spin propagation length despite DP spin
memory decay is based on the strong correlation between the
electron's displacement in space and the rotation of its spin on the
Bloch sphere. An important step toward a quantitative theory of such
correlations was made by Burkov \emph{et al.}\cite{MD} and
Mishchenko \emph{et al.},\cite{Mish} who derived equations of motion
that describe the coupling of spin and charge current degrees of
freedom in (001) GaAs QWs. Initially only the linear Rashba SO
coupling was examined, subsequently Bernevig \emph{et al.}\cite{SU2}
and Stanescu and Galitski \cite{cubic} extended the theory to
include the linear and cubic Dresselhaus terms, respectively.

The equations of motion can be solved to obtain the normal modes of
the coupled system, which are waves of mixed electrical current and
spin polarization. There exist four such modes, reflecting three
spin degrees of freedom ($S_x, S_y$, and $S_z$) and the charge
density, $n$. For wave vectors, $\mb{q}$, parallel to the directions
$[110]$ and $[1\bar{1}0]$, the four modes decouple into two pairs;
in one the spin precesses in a plane containing $\mb{q}$ and the
normal direction $\mb{\hat{z}}$, in the other the current is coupled
to the component of in-plane spin polarization perpendicular to
$\mb{q}$.

The spin precession mode is the one relevant to spin polarization
memory. For example, the decay rate of this mode at $\mb{q}=0$ is
precisely the DP decay rate, $1/\tau_s$.  In the absence of
spin-space correlation, the decay rate, $\gamma_q$, of a spin
polarization wave would increase monotonically with $q$,
\emph{i.e.}, $\gamma_q = 1/\tau_s + D_sq^2$, where $D_s$ is the spin
diffusion coefficient. Instead, it was predicted \cite{MD} that for
Rashba SO coupling the minimum decay rate occurs at nonzero wave
vector, at which point $\gamma_q$ is approximately half the DP rate.
Bernevig \emph{et al.}\cite{SU2} showed theoretically that the
minimum $\gamma_q$ is further reduced when both Rashba and linear
Dresselhaus interactions are nonzero and vanishes when the strength
of the two couplings is equal. The resulting ``persistent spin
helix" (PSH) was shown to be a conserved quantity of a newly found
SU(2) symmetry that arises when $\alpha=\beta_1$ and the cubic
Dresselhaus term ($\beta_3$) is zero.\cite{SU2} However, Stanescu
and Galitski \cite{cubic} showed that perfect SU(2) is broken when
$\beta_3\neq 0$, leading to large, but not infinite, PSH lifetime.
Recently, using the transient spin grating technique, Koralek
\emph{et al.}\cite{jake} observed the PSH mode experimentally by
independently tuning the Rashba and linear Dresselhaus couplings.

The question that arises is whether the PSH effect can be exploited
to lengthen the distance that a packet of spin polarization can
propagate in an applied electric field.  In this paper we address
this question by analyzing the effects of an in-plane electric ($E$)
field on the spin-precession modes. We focus on $\mb{E}\parallel
\mb{q}$, which is the orientation relevant to the drift of spin
polarization. To predict the spin memory length it is necessary to
determine how the applied field modifies both the real ($\Re$) and
imaginary ($\Im$) parts of the normal-mode frequency, $\omega(q)$ of
spin-polarization modes. The real part is related to the drift
velocity whereas the imaginary part is related to the lifetime. The
modification of $\Re\{\omega(q)\}$ is linear in $E$ (to lowest
order), whereas the affect of $E$ on $\Im\{\omega(q)\}$ is
quadratic.  Kleinert and Bryksin \cite{KB, KB2} recently have
treated this to problem to linear order in $E$, using quantum
kinetic theory, and obtained results for $\Re\{\omega(q)\}$.

In this work, we derive and solve equations of motion to quadratic
order in $E$ using a random walk (RW) approach that is different
from previous treatments of this problem. The advantages of our
approach are physical transparency and mathematical simplicity. We
construct a semiclassical random walk model that tracks the
electron's motion in real space and the propagation of its spin on
the Bloch sphere. In Sec. \ref{sec:rw}, we introduce the random walk
model, derive the equations of motion in the absence of an $E$
field, and solve for the spin-wave dispersion relations. We compare
the results thus obtained with the earlier quantum kinetic theory
approaches.\cite{SU2, cubic} In Sec. \ref{sec:E}, we include an
in-plane $E$ field, obtaining the equations of motion and the
dispersion relations to quadratic order. We use the dispersion
relations to analyze the motion of a spin-polarization packet in the
presence of the in-plane field, for different regimes of field
strength.  We illustrate the results by focusing on representative
SO couplings: linear Dresselhaus coupling only, the SU(2) case where
Rashba and Dresselhaus terms are equal, and the case of SU(2) broken
by a small cubic Dresselhaus term. A brief summary is given in Sec.
\ref{Summary and conclusion}.

\section{Random walk model\label{sec:rw}}

As mentioned above, as an electron propagates between scattering
events, SO coupling causes its spin to precess. Thus, as the
electron performs an RW in real space, its spin performs an RW on
the Bloch sphere. We consider a 2D electron gas with both structure
and bulk inversion asymmetry. The SO Hamiltonian for conduction band
electrons in a III-V semiconductor QW grown in the [001] direction
(taken as $\mb{\hat{z}}$-direction) is given by,
\begin{equation}
H_{SO}=\mb{\Omega}\cdot\mb{s},
\end{equation} where,
\begin{equation}
\mb{\Omega}=2k_F\left\{\mb
{\hat{x}}\left[\alpha-\beta_1-\frac{2\beta_3(v_x^2-v_y^2)}{v_F^2}\right]v_y
-\mb{\hat{y}}\left[\alpha+\beta_1-\frac{2\beta_3(v_x^2-v_y^2)}{v_F^2}\right]v_x\right\},
\end{equation}
$\mb{s}=\hbar\mb{\sigma}/2$ is the electron spin, $v_x$ and $v_y$
are the components of velocity in the [110] and [1$\bar{1}$0]
directions, $\alpha$, $\beta_1$, and $\beta_3$ are dimensionless
quantities describing the strength of the Rashba, linear, and cubic
Dresselhaus spin-orbit couplings, respectively, and $k_F$ is the
Fermi wave vector. Spins precess about the effective SO field
according to
\begin{equation}\label{times}
\frac{d\mb{s}}{dt}=\mb{\Omega}\times\mb{s}.
\end{equation}

We assume that the impurity potential is short range so that there
is no correlation between the scattering events. In the absence of
the $E$ field, electrons perform an isotropic 2D random walk with
$\mb{v}_n$ (velocity between the $n$th and $(n+1)$th scattering
events) given by $v_F\mb{\hat{t}}_n$, where
$\mb{\hat{t}}_n=(\cos\theta,\sin\theta)$ is a random two-dimensional
unit vector with a uniform probability density $p_n(\theta)=1/2\pi$.
The displacement from $n$th to $(n+1)$th step is given by
\begin{equation}
\mb{r}_{n+1}-\mb{r}_n=\mb{v}_n\tau,
\end{equation}
where $\tau$ is the electron scattering time. In the following we
consider $\Omega\tau$, the change in angle of the electron's spin
between scattering events, as a small parameter. In this case we can
obtain from Eq. (\ref{times}) the change in the spin direction
during the mean-free time as a series expansion in $\Omega\tau$,
\begin{equation}\label{spin}
\Delta
\mb{s}_n\equiv\mb{s}_{n+1}-\mb{s}_n=\mb{\Omega}_n\tau\times\mb{s}_n
+\frac{1}{2}\mb{\Omega}_n\tau\times(\mb{\Omega}_n\tau\times\mb{s}_n),
\end{equation}where we retain terms to second order.

Let $P_{n}(\mb{r})$ be the probability that after $n$ steps of
random walk the electron arrives at position $\mb{r}$ and
$D_{n}(\mb{r};\mb{s})$ be the conditional probability that given the
electron is at $\mb{r}$, its spin is $\mb{s}$. The joint probability
$P_{n}(\mb{r})D_{n}(\mb{r};\mb{s})$ satisfies the following
recursion relation:
\begin{equation}\label{recursion}
P_{n+1}(\mb{r})D_{n+1}(\mb{r};\mb{s})= \langle
P_n(\mb{r}-\mb{v}_n\tau)D_{n}(\mb{r}-\mb{v}_n\tau;\mb{s}-\Delta\mb{s}_n)\rangle,
\end{equation} where $\langle\rangle$ denotes average over $\mb{\hat{t}}_n$, i.e.,
$\langle A_n\rangle=\int_0^{2\pi}A_n(\theta)p_n(\theta)d\theta$.
Once $P_{n}(\mb{r})D_{n}(\mb{r};\mb{s})$ is determined, the
magnetization can be obtained from the following integral on the
Bloch sphere:
\begin{equation}\label{magnetization}
\mb{m}_n(\mb{r})=\int_{S^2}\mb{s}P_{n}(\mb{r})D_{n}(\mb{r};\mb{s})d\Sigma.
\end{equation}
By substituting Eq. (\ref{recursion}) into Eq.
(\ref{magnetization}), we obtain,
\begin{equation}\label{m}
\mb{m}_{n+1}(\mb{r})=\langle\int_{S^2}\mb{s}P_n(\mb{r}-\mb{v}_n\tau)
D_{n}(\mb{r}-\mb{v}_n\tau;\mb{s}-\Delta\mb{s}_n)d\Sigma\rangle.
\end{equation} Taylor series expansion on the right hand side of Eq. (\ref{m}) yields,
\begin{equation}
\begin{split}
\mb{m}_{n+1}(\mb{r})=&\langle\int_{S^2}[\mb{s}+\mb{\Omega}_n\tau\times\mb{s}+\frac{1}{2}\mb{\Omega}_n\tau\times(\mb{\Omega}_n\tau\times\mb{s})]
\{P_n(\mb{r})D_{n}(\mb{r};\mb{s})\\
&-\mb{v}_n\tau\cdot\nabla[P_n(\mb{r})D_{n}(\mb{r};\mb{s})]
+\frac{1}{2}\mb{v}_n\tau\cdot\nabla\nabla[P_n(\mb{r})D_{n}(\mb{r};\mb{s})]\cdot\mb{v}_n\tau\}d\Sigma\rangle.
\end{split}
\end{equation}
Again retaining terms to second order, we can write,
\begin{equation}\label{eq:I}
\mb {m}_{n+1}=\mb {I}_1+\mb {I}_2+\mb {I}_3,
\end{equation}where,
\begin{equation}
\begin{split}
\mb{I}_1=&\langle\int_{S^2}\mb{s}\{P_n(\mb{r})D_{n}(\mb{r};\mb{s})-\mb{v}_n\tau\cdot\nabla[P_n(\mb{r})D_{n}(\mb{r};\mb{s})]\\
&+\frac{1}{2}\mb{v}_n\tau\cdot\nabla\nabla[P_n(\mb{r})D_{n}(\mb{r};\mb{s})]\cdot\mb{v}_n\tau\}d\Sigma\rangle,
\end{split}
\end{equation}
\begin{equation}
\mb{I}_2=\langle\int_{S^2}[\mb{\Omega}_n\tau\times\mb{s}]
\{P_n(\mb{r})D_{n}(\mb{r};\mb{s})-\mb{v}_n\tau\cdot\nabla[P_n(\mb{r})D_{n}(\mb{r};\mb{s})]\}d\Sigma\rangle,
\end{equation}and
\begin{equation}
\mb{I}_3(\mb{r})=\langle\int_{S^2}[\frac{1}{2}\mb{\Omega}_n\tau\times(\mb{\Omega}_n\tau\times\mb{s})]
\{P_n(\mb{r})D_{n}(\mb{r};\mb{s})\}d\Sigma\rangle.
\end{equation}
Upon performing the average over $\mb{\hat{t}}_n$, all terms that
linear in $\mb{v}_n$ or $\mb{\Omega}_n$ vanish by symmetry, leading
to,
\begin{equation}
\mb{I}_1=\mb{m}_n+\Pi_{op}\tau^2\mb{m}_n,
\end{equation}
\begin{equation}\label{eq:I2}\begin{split}
\mb{I}_2=&-\mb{\hat{x}}\langle\Omega_{ny}v_{nx}\rangle\tau^2\frac{\partial
m_{nz}}{\partial
x}+\mb{\hat{y}}\langle\Omega_{nx}v_{ny}\rangle\tau^2\frac{\partial
m_{nz}}{\partial y}\\
&+\mb{\hat{z}}\left (\langle\Omega_{ny}v_{nx}\rangle\frac{\partial
m_{nx}}{\partial x}-\langle\Omega_{nx}v_{ny}\rangle\frac{\partial
m_{ny}}{\partial y}\right )\tau^2, \end{split}\end{equation}
\begin{equation}
\mb{I}_3=-\frac{\tau^2}{2}\left(\mb
{\hat{x}}\langle\Omega_{yn}^2\rangle m_{nx}+\mb
{\hat{y}}\langle\Omega_{xn}^2\rangle m_{ny}+\mb
{\hat{z}}\langle\Omega_{n}^2\rangle m_{nz}\right ),
\end{equation}
where
\begin{equation}
\Pi_{op}\equiv \frac{1}{2}\left(\langle v_x^2\rangle
\frac{\partial^2}{\partial x^2}+\langle v_y^2\rangle
\frac{\partial^2}{\partial y^2}\right ).
\end{equation}
Taking the continuum limit $\mb {m}_n \rightarrow \mb {m}(t)$,
$\left(\mb{m}_{n+1}-\mb{m}_{n}\right)/\tau\rightarrow d\mb{m}/dt$,
and substituting into Eq. (\ref{eq:I}), we obtain the equation of
motion for the magnetization vector.  Resolving the vector equation
into components yields three scalar equations,
\begin{equation}
\frac{1}{\tau}\frac{\partial m_x}{\partial t}=\Pi_{op}
m_x-\frac{1}{2}\langle\Omega_{y}^2\rangle m_{x}-\langle\Omega_y
v_x\rangle\frac{\partial m_z}{\partial x},
\end{equation}

\begin{equation}
\frac{1}{\tau}\frac{\partial m_y}{\partial t}=\Pi_{op}
m_y-\frac{1}{2}\langle\Omega_{x}^2\rangle m_{y}+\langle\Omega_x
v_y\rangle\frac{\partial m_z}{\partial y},
\end{equation}
\begin{equation}
\frac{1}{\tau}\frac{\partial m_z}{\partial t}=\Pi_{op}
m_z-\frac{1}{2}\langle\Omega^2\rangle m_{z}+\langle\Omega_y
v_x\rangle\frac{\partial m_x}{\partial x}-\langle\Omega_x
v_y\rangle\frac{\partial m_y}{\partial y}.
\end{equation}
Solving the equations of motion for eigenmodes with wave vector
parallel to $\mb{\hat{x}}$ yields the dispersion relation,
\begin{equation}\label{dispersion1}
\frac{i\omega_{\pm}(q)}{\tau}=\frac{1}{4}\left
(2\langle\Omega^2\rangle-\langle\Omega_x^2\rangle\right)+\frac{1}{2}\langle
v_x^2\rangle
q^2\pm\sqrt{\frac{\langle\Omega_x^2\rangle^2}{16}+q^2\langle\Omega_y
v_x\rangle^2 }.
\end{equation}
This dispersion relation corresponds to modes in which the spin
polarization spirals in the $x$-$z$ plane. Note that $\omega(q)$ is
purely imaginary so that for all wave vectors the spin-polarization
wave decays exponentially with time. However, the dispersion
relation differs from ordinary diffusion, where $i\omega \propto
1/\tau+ Dq^2$. The difference can be traced to the terms in Eq.
(\ref{eq:I2}) that are proportional to the first derivative of spin
density with respect to position - these terms are absent in the
usual diffusion equation.  The coefficients of these additional
terms are the cross-correlation functions, $\left\langle \Omega_x
v_y\right\rangle$ and $\left\langle \Omega_y v_x\right\rangle$,
which shows explicitly that the anomalous diffusion is a consequence
of the correlation between the electron's motion in real space and
the propagation of its spin on the Bloch sphere.

In the SU(2) case ($\alpha=\beta_1$ and $\beta_3=0$), Eq. (\ref
{dispersion1}) simplifies to,
\begin{equation}
i\omega_{\pm}(q)=\frac{1}{4} v_F^2 \tau\left ( q\pm q_0 \right
)^2\equiv D(q\pm q_0)^2,
\end{equation}
where $D\equiv v_F^2\tau/4$ and $q_0\equiv 4k_F\beta_1$. The
vanishing decay rate of the $\omega_-$ mode at $q=q_0$ indicates the
appearance of a conserved quantity - a helical spin-polarization
wave or ``persistent spin helix".\cite{SU2}

The dispersion relations obtained above for the spiral polarization
waves are the same as those obtained previously, including the cubic
Dresselhaus term.\cite{SU2, cubic} We note, however, that while the
RW approach accurately describes the spiral coupling of $x$-$z$
components of spin, it does not capture the coupling between charge
current and the $y$ component of spin that appears in the quantum
kinetic formulation. This is because the RW approach does not
include relaxation to the equilibrium state.  In other words,
between consecutive scattering events the electron's spin precesses
about $\mb{b}(\mb{k})$ but has no tendency to spiral in toward it.
Thus the well-known current-induced spin polarization (CISP) effect
\cite{Mish} is not predicted. To recover CISP requires adding to Eq.
(\ref{times}) a phenomenological Gilbert damping term,
\begin{equation}
\frac{d\mb{s}}{dt}=\lambda_G
\mb{s}\times\left(\mb{\Omega}\times\mb{s}\right),
\end{equation}where $\lambda_G$ is the damping parameter.

\section{Spin Helix dynamics in the presence of an electric
field}\label{sec:E}

In this section, we explore how the spin dynamics change in the
presence of an $E$ field parallel to the wave vector of the spin
spiral. To include the effect of $E$ we add a drift term to the
velocity at each random walk step,
\begin{equation}
\mb{v}_n=v_F\mb{\hat{t}}_n+v_d\mb{\hat{x}},
\end{equation}
where $v_d$ is the drift velocity assumed to be a linear function of
$E$. We assume further that the electric field does not change the
shape of the impurity potential and therefore the scattering
probability density is still uniform.

The drift velocity modifies the precession vector, adding a fixed
precession
\begin{equation}
\mb{\Omega}_d\equiv
-2\mb{\hat{y}}k_F\left[\alpha+\beta_1-\frac{2\beta_3(v_x^2-v_y^2)}{v_F^2}\right]v_d,
\end{equation} to $\mb{\Omega}_n$ at each step of the random walk. Substituting and
following the same strategy as before, we obtain,
\begin{equation}
\mb{I}_1(\mb{E})=\mb{I}_1-v_d\tau\frac{\partial \mb{m}}{\partial x},
\end{equation}
\begin{equation}
\mb{I}_2(\mb{E})=\mb{I}_2+\mb{\Omega}_d\tau \times \mb {m},
\end{equation}
\begin{equation}
\mb{I}_3(\mb{E})=\mb{I}_3,
\end{equation}
where the $\mb{I}_{1,2,3}(\mb{E})$ are the quantities
$\mb{I}_{1,2,3}$ evaluated in the presence of the electric field.
The field alters the equations of motion in two ways. First, new
terms appear that are linear in $E$.  The new term added to
$\mb{I}_1$ converts the time derivative of $\mb{m}$ to the
convective derivative, that is the time derivative in a frame moving
with the drifting electrons.  The term added to $\mb{I}_2$ indicates
that the $E$ field introduces uniform precession about the
$\mb{\hat{y}}$ axis, when viewed in the frame co-moving with
$\mb{v}_d$. The second type of modification is quadratic in $E$; the
field increases $\langle\Omega_y^2\rangle$ by the additive factor
$\Omega_d^2$ and the mean-square velocity $\langle v_x^2 \rangle$ by
the factor $\langle v_d^2 \rangle$.

Solving for normal modes with wave vector parallel to
$\mb{\hat{x}}$, we obtain
\begin{equation}\label{io0}
i\omega_{\pm}(q)= \frac{1}{4}\left
(2\langle\Omega^2\rangle-\langle\Omega_x^2\rangle\right)\tau+\frac{1}{2}\langle
v_x^2\rangle\tau
q^2+iv_dq\pm\sqrt{\frac{\langle\Omega_x^2\rangle^2\tau^2}{16}+\left
( q\langle\Omega_y v_x\rangle\tau +i\Omega_d\right )^2 }.
\end{equation}
To linear order in $E$, this dispersion relation is the same as that
obtained by Kleinert and Bryksin.\cite{KB, KB2} In the presence of
the electric field $\omega(q)$ acquires a real part, which describes
the propagation of spin polarization. Equation (\ref{io0}) also
describes the modifications of the spin polarization lifetime that
appear at second order in $E$. In the following we discuss the spin
dynamics that emerge from this dispersion relation for
representative SO Hamiltonians.

\subsection{SU(2) case}\label{sec:SU(2)}

\begin{figure}
    \centering
    \subfigure[]{
      \label{fig:ratePSH}
      \includegraphics[width=0.45\textwidth]{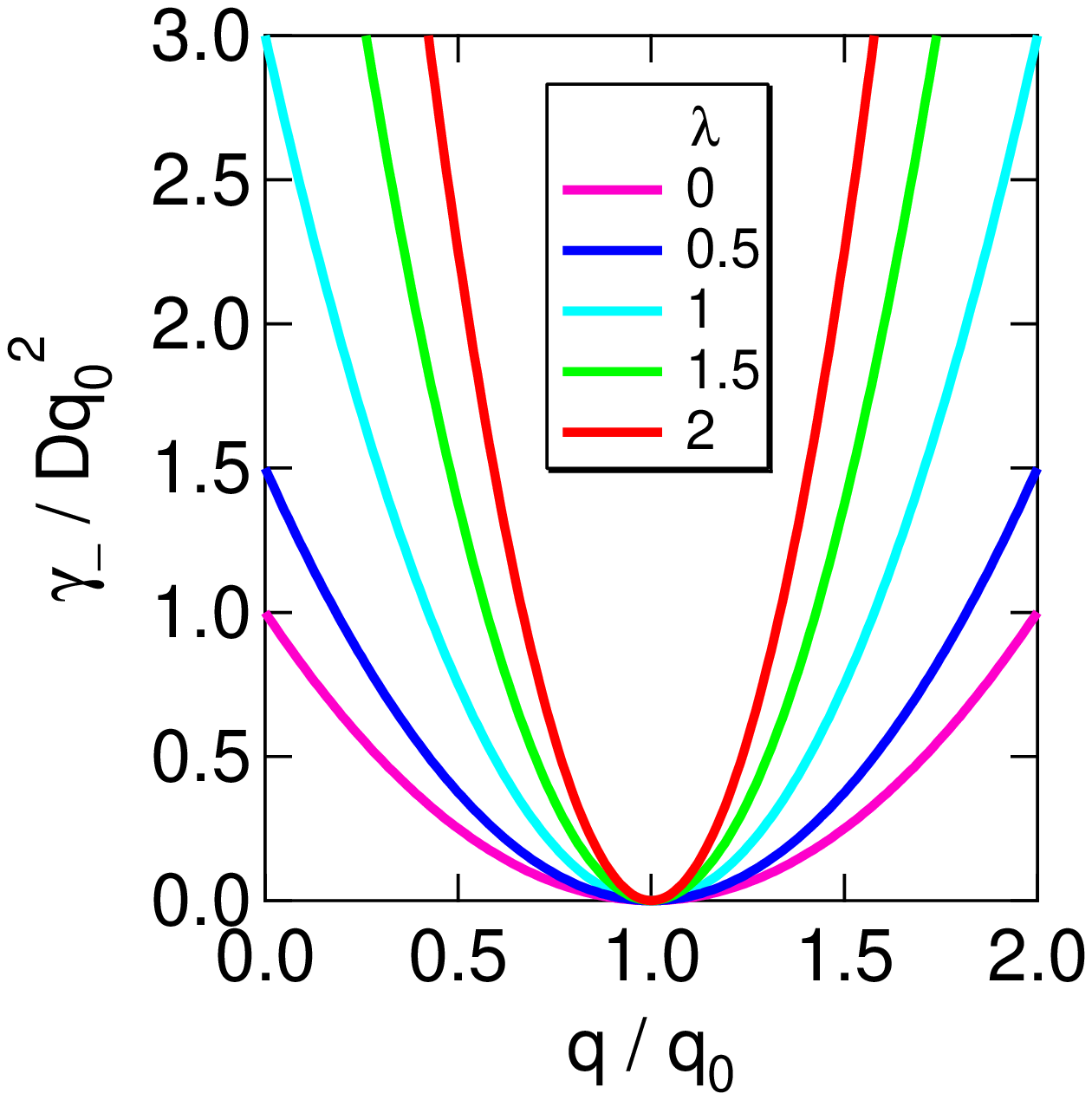}}
    \subfigure[]{
      \label{fig:phasePSH}
      \includegraphics[width=0.45\textwidth]{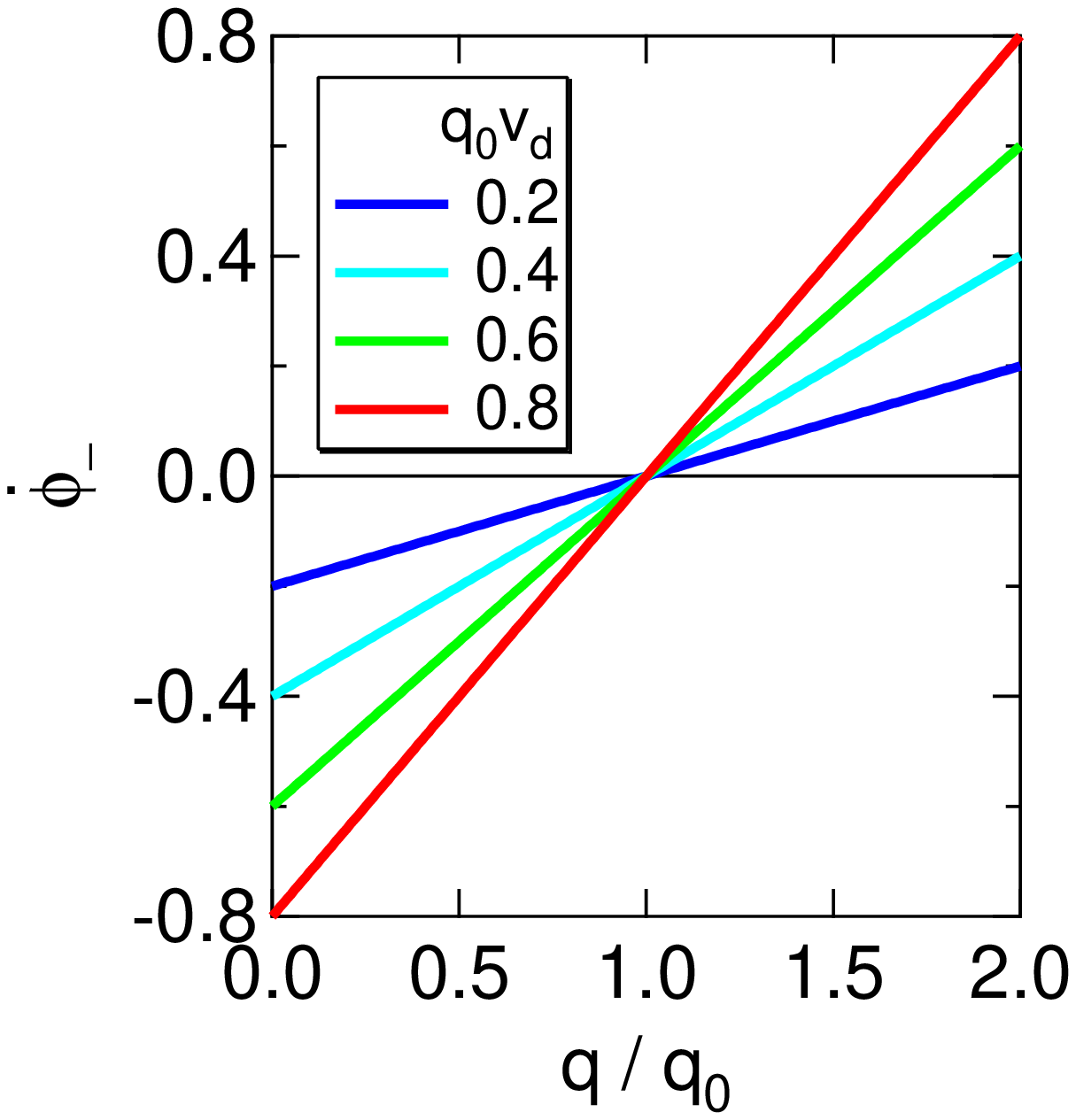}}
           \caption{(Color online) The dispersion relations for (a) the decay rate and (b) the rate of phase change
of the SO enhanced mode in the SU(2) case. (a) The decay rate
$\gamma_-(q)$ increases with the drift velocity ($\lambda\equiv
v_d/v_F$) but always vanishes at the resonant wave vector $q_0$. (b)
The rate of phase change $\dot{\phi}_-(q)$ is proportional to the
drift velocity $v_d$ and it crosses zero at the resonant wave vector
$q_0$.}\label{fig:PSH}
\end{figure}

For the case of $\alpha=\beta_1, \beta_3=0$, the dispersion relation
simplifies to,
\begin{equation}
i\omega_\pm(q)=D\left(1+2\lambda^2\right)\left(q\pm
q_0\right)^2+iv_d\left(q\pm q_0\right),
\end{equation}
where $\lambda\equiv v_d/v_F$. To distinguish the lifetime and
propagation effects we write the dispersion relation in the form,
\begin{equation}
i\omega(q)=\gamma(q)+i\dot{\phi}(q),
\end{equation}
where $\gamma(q)$ is the decay rate and $\dot{\phi}(q)$ is the rate
of phase advance. The real and imaginary parts of $i\omega_-(q)$,
corresponding to the longer lived of the two modes, are plotted in
Fig. \ref{fig:PSH}. As is apparent from Fig. \ref{fig:ratePSH}, the
spin polarization lifetime, $1/\gamma_-(q)$ remains infinite at the
PSH wave vector, despite the presence of the electric field. This
result is consistent with the theoretical prediction that at the
SU(2) point the spin helix generation operators commute with all
perturbation terms that are not explicitly spin dependent.\cite{SU2}
However, the field increases the effective diffusion constant by the
factor $\lambda^2$ so that the decay rate for $q\neq q_0$ increases
rapidly when the drift velocity approaches the thermal velocity of
the electrons. The spin helix generation operators won't commute
with the Hamiltonion if there exists a spatial disorder of SO
interactions.\cite{Sherman1, Sherman2}

The rate of phase advance [plotted in Fig. \ref{fig:phasePSH}]
vanishes at $q=q_0$, i.e., the PSH is stationary, despite the fact
that the Fermi sea of electrons is moving by with average velocity
$\mb{v}_d$. Moreover, spin spirals with $q<q_0$ will appear to move
backward, that is, opposite to the direction of electron flow.
Although unusual, this property can be understood by considering the
spin dynamics in a frame moving with velocity $\mb{v}_d$. In this
frame $\mb{E}$ parallel to $\mb{\hat{x}}$ is perceived as a
precession vector $\mb{\Omega}_d=-4\beta_1 v_d\mb{\hat{y}}=-v_d q_0
\mb{\hat{y}}$. Therefore in the moving frame $\phi_\pm(x',t')=\pm
qx'-v_d q_0 t$. Transforming back to the laboratory frame then
yields $\dot{\phi}_\pm(q)=v_d (q\pm q_0)$.

\begin{figure}
      \includegraphics[width=0.6\textwidth]{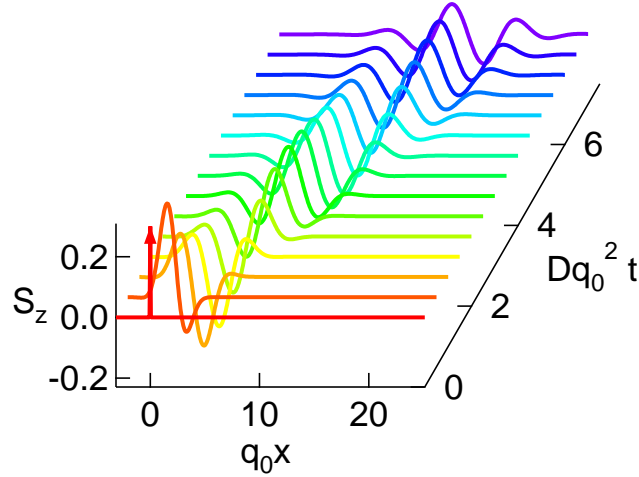}
      \caption{(Color online) The space-time evolution of $S_z$ with a normalized
       $\delta$-function injection at $x=0,t=0$, and drift velocity $v_d=2Dq_0$ in the SU(2) case.
       The spin polarization develops into a conserved stationary wave with a Gaussian wave packet.}
      \label{fig:waterPSH}
\end{figure}

The nature of spin propagation at the SU(2) symmetry point can be
made more clear if we Fourier transform from the wave vector to
spatial domain.  If we inject a $\delta$-function stripe of $z$
polarized spins at $x=0$, the space-time evolution of $S_z$ is
proportional to the propagator, $G_z(x,t)$, where
\begin{equation}\label{FT}
G_z(x,t)\propto\int dq e^{iqx}\left(A_+e^{-i\omega_+
t}+A_-e^{-i\omega_- t}\right ),
\end{equation}
where $A_+$ and $A_-$ are the weighting factors for the passive and
active modes, respectively and $A_+=A_-=1/2$ in the SU(2) case. Upon
substituting the dispersion relations $\omega_\pm(q)$, we obtain,
\begin{equation}\label{Green}
G_z(x,t)\propto \frac{1}{\sqrt{Dt}}\cos(q_0
x)\exp{\left[-\frac{(x-v_dt)^2}{4Dt}\right]}.
\end{equation}
The spin propagator is the product of a Gaussian envelope function
and a static spin wave with wave vector $q_0$.  The envelope
function is the one-dimensional diffusion propagator with width
proportional to $\sqrt{Dt}$ and drift velocity $v_d$. An
illustration of the space-time evolution described by this
propagator is provided Fig. \ref{fig:waterPSH}, for a drift velocity
$v_d=2Dq_0$. Note that the phase of the spin wave modulated by the
Gaussian envelope remains stationary as the packet drifts and
diffuses. This contrasts with the more familiar wave packet, where
the modulated wave and envelope functions both propagate, albeit
with velocities that may differ.

\subsection{SU(2) broken by cubic Dresselhaus term}\label{sec:SU(2)beta3}

When SU(2) is exact, the integral of the Gaussian envelope function
is conserved, even in the presence of an $E$ field.  However,
Stanescu and Galitski \cite{cubic} have shown theoretically that
$\beta_3$, which is nonzero in real systems, breaks SU(2). Koralek
\emph{et al.} \cite{jake} verified experimentally that $\beta_3$ is
indeed the factor that limits PSH lifetime in experiments on (001)
GaAs quantum wells. In this section we calculate the dispersion
relation and spin packet time evolution in the presence of a small
cubic Dresselhaus term.

It was shown previously that when $\beta_3$ is small, the maximum
lifetime occurs when the Rashba interaction $\alpha=\beta_1-\beta_3$
(Ref. \onlinecite{cubic}).  We consider a QW with Rashba coupling
tuned to this value and assume that $\beta_3\ll\beta_1$. This
condition is met in QWs in the 2D limit, where $k_F d\ll 1$ ($d$ is
the well width). In this case the dispersion relation in the
presence of the electric field can be written as
\begin{equation}
i\omega_\pm (q)\cong 6Dk_F^2\beta_3^2+ D \left(q\pm
q_0\right)^2+iv_d \left(q\pm q_0\right)\mp iv_d\Delta q,
\end{equation}where $q_0\equiv 4k_F(\beta_1-\beta_3)$ and $\Delta q=2k_F \beta_3$.
Performing the Fourier transform to obtain the space-time evolution
of a spin packet, we obtain,
\begin{equation}
G_z(x,t)\propto \frac{1}{\sqrt{Dt}}e^{-6Dk_F^2\beta_3^2t}\cos(q_0
x-v_d\Delta q t)\exp{\left[-\frac{(x-v_dt)^2}{4Dt}\right]}.
\end{equation}In the presence of the cubic Dresselhaus interaction the integral of the
Gaussian envelope is no longer conserved. The decay rate can be
written in the form,
\begin{equation}
\gamma=\frac{3}{8}Dq_0^2\left(\frac{\beta_3}{\beta_1}\right)^2,
\end{equation}
illustrating that although the decay rate is nonzero, it is reduced
relative to the DP relaxation rate by a factor
$\approx(\beta_3/\beta_1)^2$. This ratio is expected
theoretically,\cite{Winkler} and has been verified
experimentally,\cite{jake} to be determined by the relation,
\begin{equation}
\frac{\beta_3}{\beta_1}=\frac{k_F^2 d^2}{4\pi^2}.
\end{equation}
For quite reasonable QW parameters a $\beta_3$ to $\beta_1$ ratio of
1:100 can be achieved, equivalent to a lifetime enhancement relative
to the DP spin memory time on the order of $10^4$.

\subsection{Linear Dresselhaus coupling}\label{Dresselhaus coupling}
\begin{figure}
    \centering
    \subfigure[]{
      \label{fig:rateD}
      \includegraphics[width=0.45\textwidth]{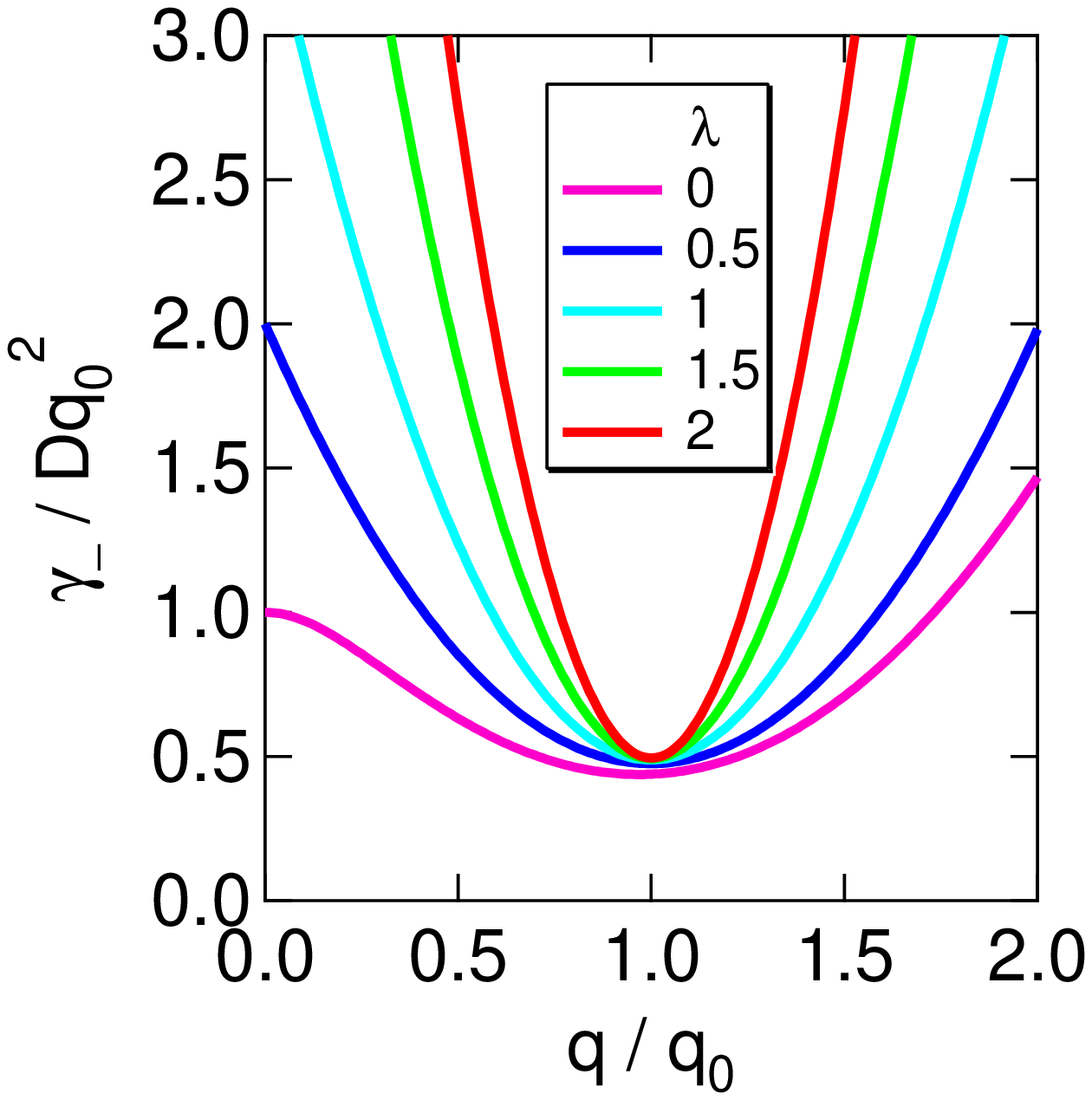}}
    \subfigure[]{
      \label{fig:phaseD}
      \includegraphics[width=0.45\textwidth]{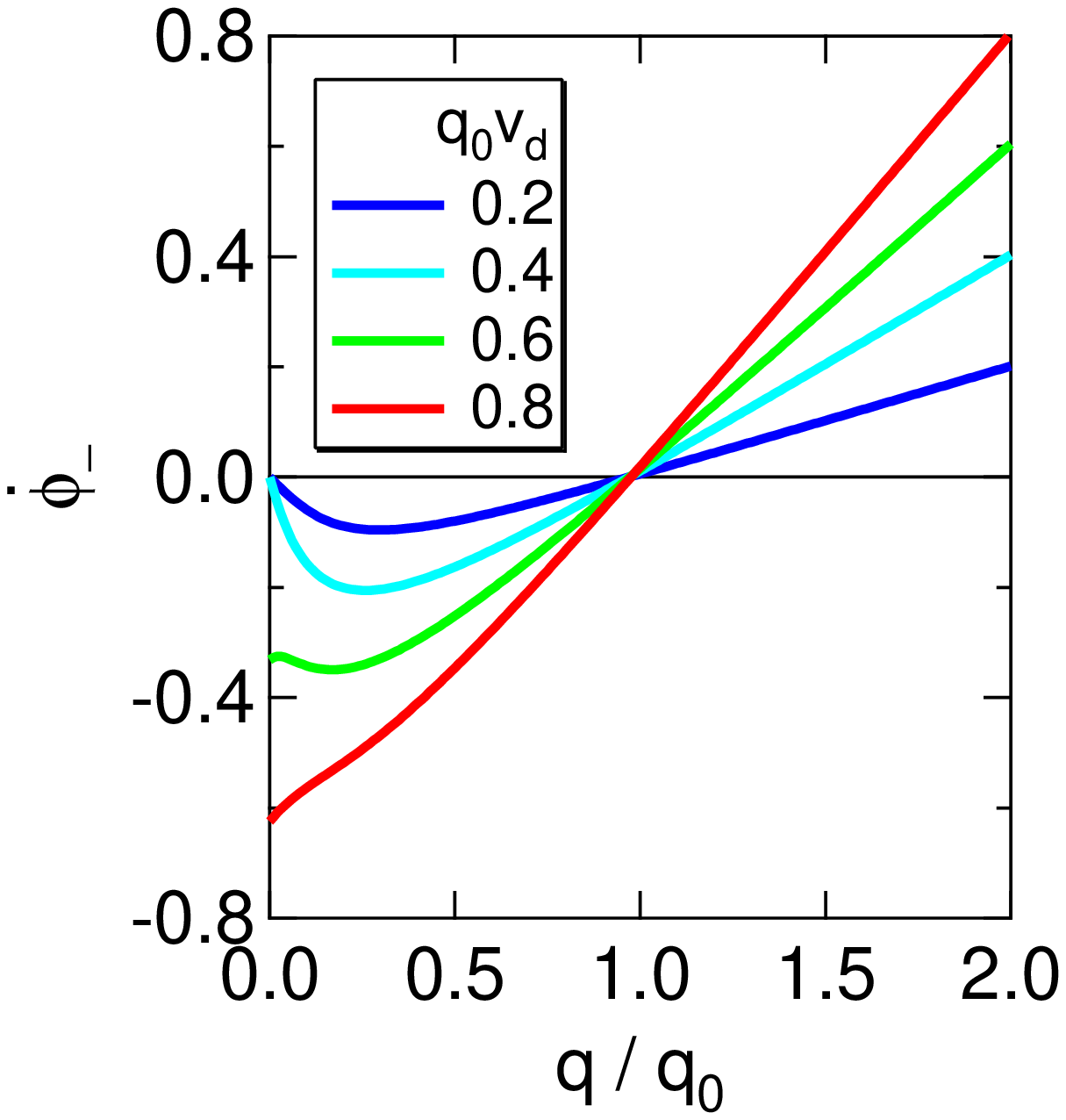}}
            \caption{(Color online) The dispersion relations for (a) the decay rate and
            (b) the rate of phase change of the SO enhanced mode in the linear-Dresselhaus-only case.
        The main features resemble those in the SU(2) case,
        both $\gamma_-(q)$ show a minimum and $\dot{\phi}_-(q)$ vanishes at $q_0$,
        but the lifetime is finite in this case.
        }\label{fig:Dresselhauss}
\end{figure}

Finally, we consider a fully symmetric well in which only the linear
Dresselhaus coupling exists. To make comparison with the SU(2)
situation, we set the strength of the linear Dresselhaus coupling be
$2\beta_1$, so that the resonant wave vector is at $q\simeq
q_0=4k_F\beta_1$. The dispersion relations $\gamma_-(q)$ and
$\dot{\phi}_-(q)$ obtained by substituting $\alpha=\beta_3=0$ and
replacing $\beta_1$ by $2\beta_1$ in Eq. (\ref{io0}) are plotted in
Fig. \ref{fig:Dresselhauss}. Some qualitative features of the
dispersion relations are similar to the SU(2) case, in that
$\gamma_-(q)$ has a global minimum and $\dot{\phi}_-(q)$ crosses
zero at $q\simeq q_0$. The most important difference is that the
minimum $\gamma_-(q)$ does not reach zero, and therefore the spin
spiral does decay. In the limit of low electric field, the lifetime
of the spin spiral is only about a factor of 2 longer than the $q=0$
(DP) lifetime.

\begin{figure}
      \includegraphics[width=0.6\textwidth]{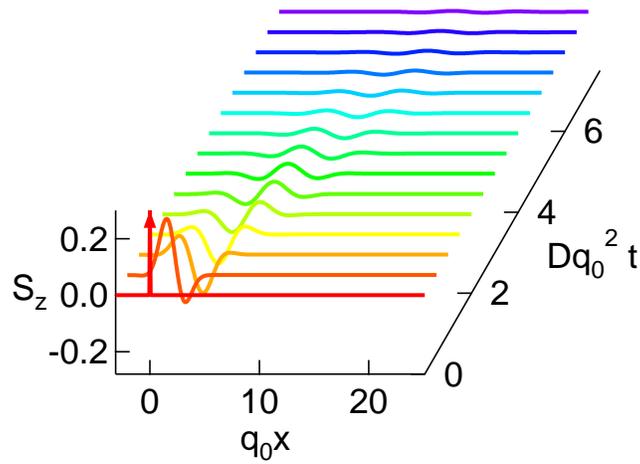}
      \caption{(Color online) The space-time evolution of $S_z$  in the linear-Dresselhaus-only case
      with the same initial condition and applied $E$ field as in the SU(2) case.
        The features are similar to those in the SU(2) case,
         except the envelope function decays exponentially.}
      \label{fig:waterDresselhaus}
\end{figure}

\begin{figure}
      \includegraphics[width=0.45\textwidth]{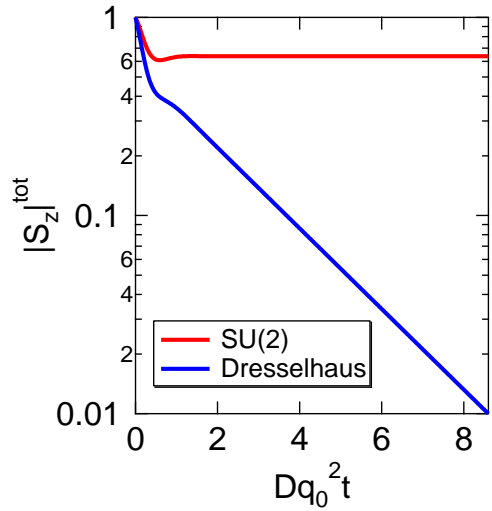}
      \caption{(Color online) The the absolute value of the spin polarization
      integrated over position as a function of time.
       In the SU(2) case, $|S_z|^{tot}$ is conserved after an initial decay;
       while in the linear-Dresselhaus-only case, $|S_z|^{tot}$ decays exponentially.}
      \label{fig:area}
\end{figure}

The propagation of a spin packet in the linear-Dresselhaus-only case
is illustrated in Fig. \ref{fig:waterDresselhaus}, using the same
initial condition and drift velocity as in SU(2) case. We performed
numerical integration of Eq. (\ref{FT}) to obtain the propagator. As
we have seen previously, a drifting and diffusing envelope function
modulates a spiral spin wave. However, now the spiral spin fades
very quickly. The contrast between linear Dresselhaus only and SU(2)
is illustrated in Fig. \ref{fig:area}, which is a plot of the
integral of the envelope as a function of time. After a rapid
initial decay, the integral is constant in the SU(2) case, whereas
with only the linear Dresselhaus interaction the integrated
amplitude decays exponentially with rate $\simeq Dq_0^2$.

\begin{figure}
    \centering
    \subfigure[]{
      \label{fig:Sxt}
      \includegraphics[width=0.45\textwidth]{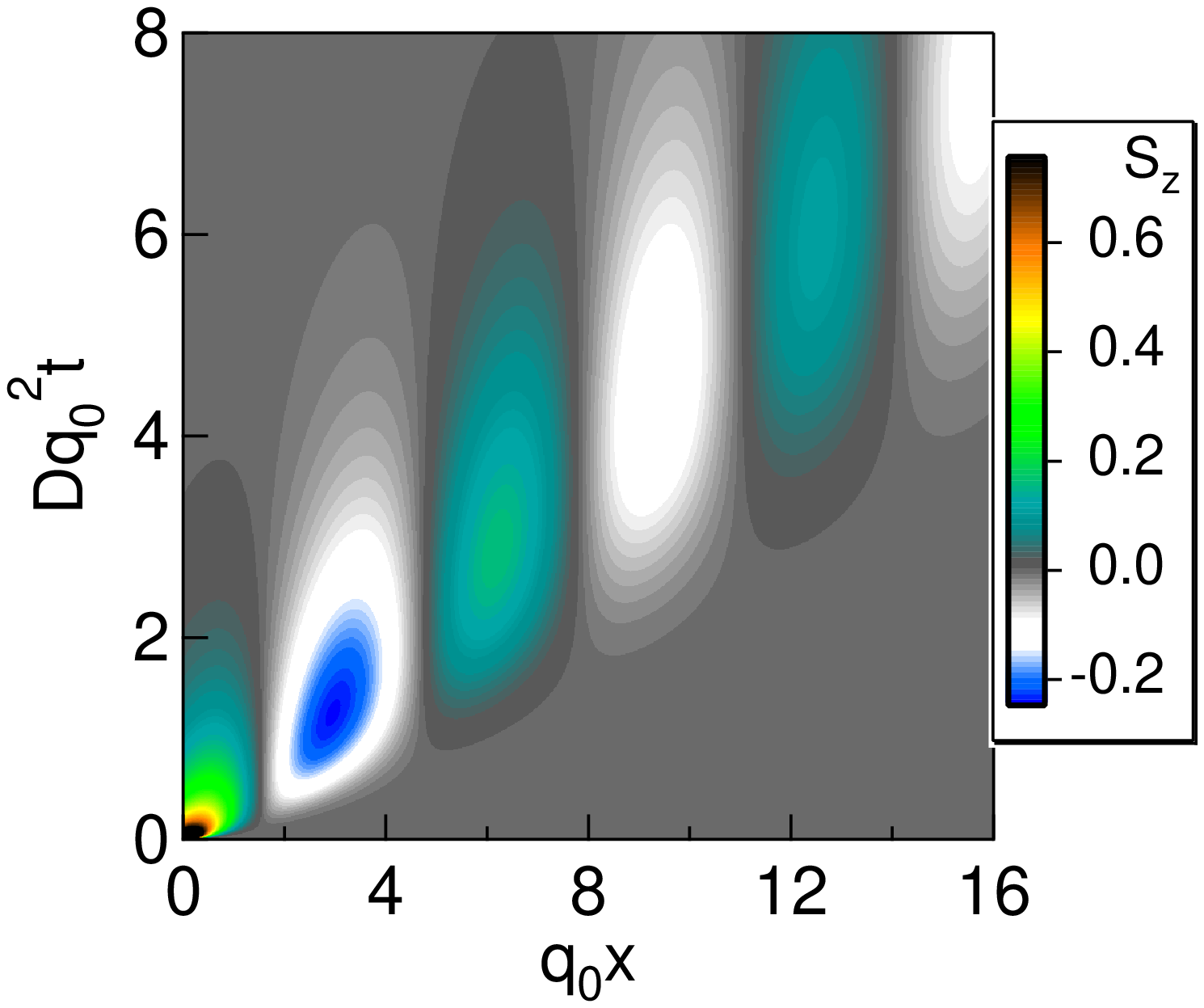}}
    \subfigure[]{
      \label{fig:Dxt}
      \includegraphics[width=0.45\textwidth]{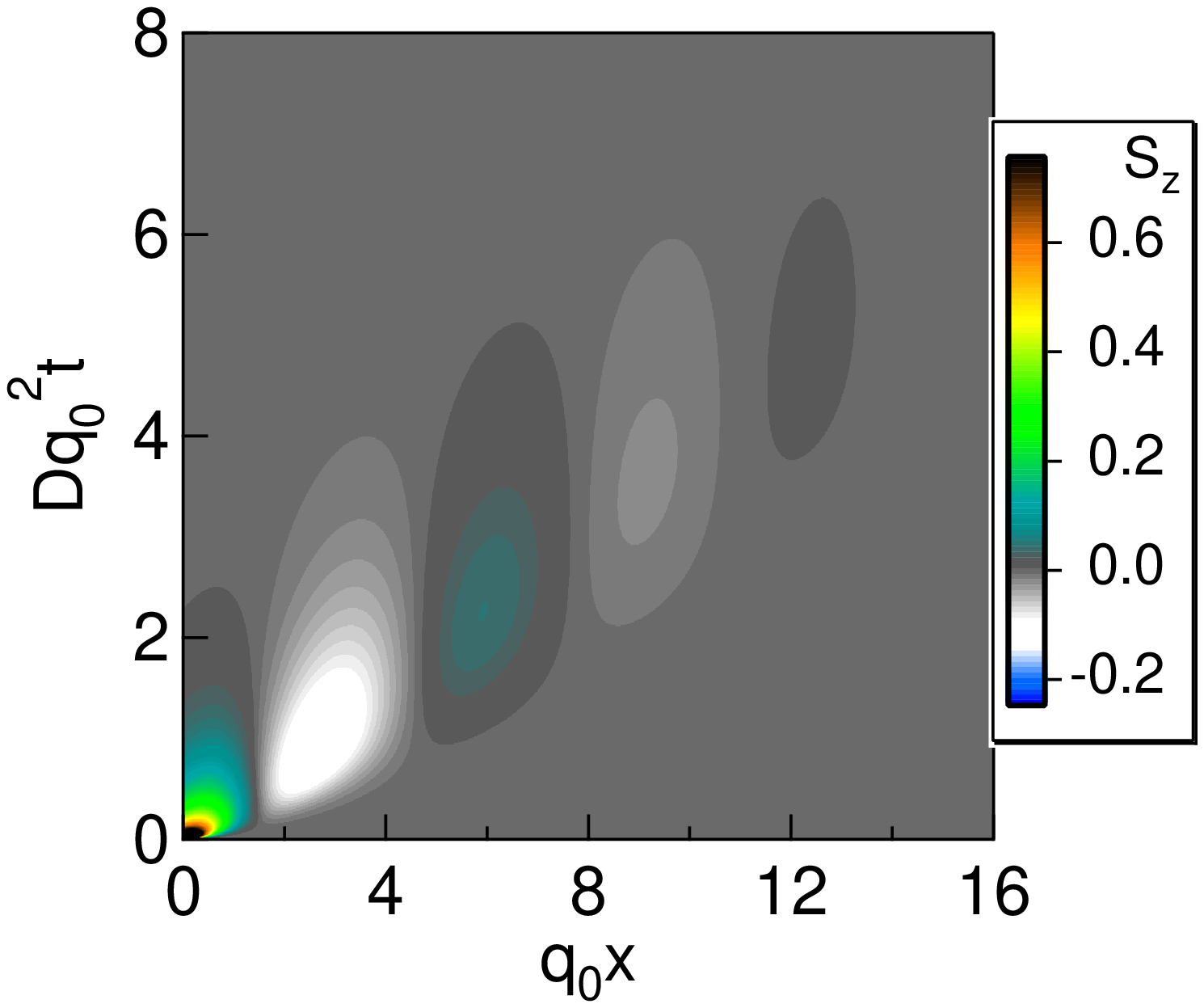}}
      \caption{(Color online) The space-time images of the spin polarization
       in the (a) SU(2) and (b) linear-Dresselhaus-only cases, respectively.}\label{fig:xt}
\end{figure}

Figure \ref{fig:xt} presents another way of visualizing the
difference in propagation for the SU(2) [Fig. \ref{fig:Sxt}] and
linear-Dresselhaus-only [Fig. \ref{fig:Dxt}] Hamiltonians.  The $z$
component of spin polarization is shown (with color coded amplitude)
as a function of time on the vertical axis and position on the
horizontal  axis.  It is clear, from the vertical orientation of the
contours that the positions of the nodes and antinodes of $S_z$ are
fixed in space.

\section{Summary and conclusion}\label{Summary and conclusion}

We have developed a random walk model to describe the time evolution
of electron spin in two dimensions in the presence of Rashba and
Dresselhaus interactions. From the random walk model we derived
equations of motion for spin polarization and obtained dispersion
relations for $\mb{q}$ parallel to one of the symmetry directions of
the Rashba/Dresselhaus Hamiltonian.  In Sec. \ref{sec:rw}, we showed
that the dispersion relations for spin-polarization waves that
spiral in the plane containing the surface normal and the wave
vector are identical to those obtained from previous
analyses.\cite{SU2, cubic} The random walk approach is instructive
in showing, in a simple but explicit way, how anomalous spin
diffusion and the persistent spin helix arise from nonvanishing
correlations between the velocity and spin precession vectors.

In Sec. \ref{sec:E}, we obtained dispersion relations for
spin-polarization waves that include the effects of an electric
field parallel to $\mb{q}$, to second order in $E$. The  terms
linear in $E$ are equivalent to those obtained from the quantum
kinetic approach.\cite{KB, KB2} To first order in $E$, the field
introduces a precession vector in the plane of the 2DEG and
perpendicular to $\mb{E}$. The precession about the $y$ axis gives
rise to an unusual behavior in that the spiral with wave vector
$q_0$ is stationary in space despite the motion of electrons in the
field; waves with $q>q_0$ propagate in the same direction as the
drifting electrons while those with $q<q_0$ propagate ``backward."
The terms that are second order in $E$ affect the decay rate of spin
polarization without changing the velocity.  The solutions obtained
when these terms are included point to the special properties of
waves with wave vector $q_0$, whose lifetime turns out to be
unchanged by the field. However, the decay rate of the all other
waves increases, in proportion to $(q-q_0)^2$.

We illustrated these results by considering three representative
spin-orbit Hamiltonians: SU(2) symmetric or $\alpha=\beta_1$ and
$\beta_3=0$; SU(2) broken by a small but nonzero $\beta_3$; and
linear Dresselhaus coupling only or $\alpha=\beta_3=0$. In order to
show the nature of spin propagation more clearly, we Fourier
transformed the solutions from wave vector to real space and
obtained the dynamics of spin-polarization packets. In all cases the
spin packets move at the electron drift velocity. In the SU(2) case
the integrated amplitude of the spin spiral is conserved, while in
the linear-Dresselhaus-only case the amplitude decays with a rate
$\sim D q_0^2$.  When SU(2) is weakly broken by small, but nonzero
$\beta_3$, the integrated amplitude decays at a rate
$\sim(\beta_3/\beta_1)^2 Dq_0^2$.

The conclusions reached by our analysis of the RW model are
consistent with a recent Monte Carlo study of a specific 2DEG
system, a (001) $\hbox{In}_{1-x}\hbox{Ga}_{x}\hbox{As}$ quantum well
with carrier density $\sim 10^{12}$ cm$^{-2}$ (Ref.
\onlinecite{Ohno}). In this study spin polarization dynamics were
calculated under conditions of steady state injection from a
ferromagnetic contact. For $\alpha/\beta_1$ ratios that are close to
unity, the spin polarization is conserved over several wavelengths
of the PSH, despite the fact that transport takes place in the
diffusive regime. Moreover, the polarization is not diminished with
increasing electric field.  The authors point out that the PSH
effect can be used to achieve a novel variation of the Datta-Das
spin-field-effect transistor (Ref. \onlinecite{DD}) in which a gate
electrode modulates the $\alpha$ to $\beta_1$ ratio only slightly
away from unity.  This has the effect of varying the wavelength of
the PSH without significantly reducing its lifetime. Thus small
changes in gate voltage can in principle lead to large changes in
source to drain conductance. Whether such a device can actually be
realized depends on two factors: fabricating ferromagnetic injectors
and analyzers with high figures of merit, and demonstrating that the
PSH effects that have been observed at temperatures below $\sim$ 100
K (Ref. \onlinecite{jake}) can be realized at room temperature.

\begin{acknowledgments}
This work was supported by the Director, Office of Science, Office
of Basic Energy Sciences, Materials Sciences and Engineering
Division, of the U.S. Department of Energy under Contract No.
DE-AC02-05CH11231.
\end{acknowledgments}

\end{document}